# Performance of PPN Guided Missile with Time-Varying Speed against Stationary Target – New Findings

Liu Yuanhe[1,2], Li Kebo[1,2], Liang Yan'gang[1,2]

1. College of Aerospace Science and Engineering, National University of Defense Technology, Changsha, 410072, China

2. Hunan Key Laboratory of Intelligent Planning and Simulation for Aerospace Missions, Changsha, 410072, China

**Abstract:** In this paper, some new results on time-varying missile against a stationary target using pure proportional navigation (PPN) are developed in the planar interception problem. First, the relative motion equation is established in arc-length domain based on the differential geometry theory, which eliminates the influence of time-varying missile speed. Then, the closed-form solution of time-varying speed missile intercepting stationary target with PPN is deduced, and the interception performance is analyzed. Additionally, considering the missile maneuvering acceleration limit, the capture region of time-varying speed missile is analyzed. Finally, the results derived in this paper are verified by numerical simulation analysis for various scenarios.

## 1 Introduction

Proportional navigation (PN) is a widely used classical guidance scheme which is mainly divided into two research directions. One is (PPN) and its extended forms, whose command acceleration is perpendicular to the velocity direction. The other is true proportional navigation (TPN) and its extended forms with command acceleration perpendicular to the line-of-sight (LOS) direction. Most of the existing missile PNs assume that the missile flight speed or relative closing speed is constant or approximately constant, but in the practical application of PN, the real-time speed is considered. In fact, many seekers use Doppler radar, which can continuously provide information about the current speed. Therefore, the method of analyzing PN guidance law is different from that of realizing PN guidance law.

In Ref. [1], the capture area and performance of PPN against stationary target under ideal conditions are studied qualitatively for the first time, and the general condition of bounded LOS rate is given. On this basis, the closed solution of true proportional navigation (TPN) attacking stationary target is derived in Ref. [2], and the necessary conditions for the capture region are given. Compared with PPN, the capture region is smaller. In addition, in Ref. [3], the closed-form solution of three-point guidance law tracking non-maneuvering target is derived, and the analytical solutions of interception flight time, speed increment, and initial capture conditions are obtained.

The results are extended to maneuverable targets, providing a possible unified analysis method for the study of various guidance laws in Ref. [4]. Besides, the closed-form solution of PPN intercepting non-maneuverable targets is extended in Ref. [5], which can be expressed by the infinite product with uniform fast convergence, and analyzes the necessary and sufficient conditions for the change of LOS rate. The generalized proportional navigation (GPN) with a certain angle between the guidance acceleration and the normal direction of LOS is defined in Ref. [6]. In addition, the closed-form solution of GPN against non-maneuverable targets is deduced, which shows that the deflection angle is related to the guidance acceleration command and interception time. The unified form of proportional navigation is introduced in Ref. [7] further, in which the closed-form solution analysis method is extended to a more general form. Then the explicit expressions of capture region, guidance acceleration, energy consumption, and flight time in the general form are derived. Finally, the closed-form solutions of several guidance laws are given through this general method. With that, the closed-form solution of GPN to maneuvering target is deeply analyzed, and the influence of deviation angle of command acceleration on capture region and energy consumption are studied in Ref. [8]. The results show that the maneuvering form of the target reduces capture region and increases energy consumption. Ref. [9] proposed an ideal proportional navigation (IPN) with the command acceleration perpendicular to the relative velocity direction, and derived the closed-form solution of IPN. The analysis shows that the IPN has a larger capture region without considering the energy consumption.

The research in Ref. [3]~[9] is based on the assumption that the speed is not disturbed, that is, the missile speed is assumed to be constant. Many results are not necessarily applicable to the case of arbitrary time-varying missile speed while taking reality into consideration. At present, there are few studies on arbitrary variation of missile speed. The capture region of real true proportional navigation (RTPN) for non-maneuvering targets is studied and the capture equation is derived and analyzed qualitatively in Ref. [10] by eliminating the relative speed, which relaxes the constant-speed assumption. The PPN guidance performance and the capture region of missile with time-varying speed against stationary target is explored in Ref. [11] preliminarily. In this paper, the arc-length domain is introduced to eliminate the influence of arbitrary time-varying speed assumption for the first time, which is the prototype of classical differential geometric curve theory.

The classical differential geometric curve theory is to design and analyze the guidance law in the arc-length domain, which can eliminate the influence of the missile time-varying speed. The guidance command of tactical missile guidance against maneuvering target is analyzed by using the classical differential geometric curve theory in Ref. [12] and [13]. On this basis, a new differential geometric guidance law is designed and the closed-form solutions of LOS rate and capture region are derived in arc-length domain by analyzing the characteristics of pointing velocity vector of virtual missile through Frenet-Serret formula. However, the missile velocity is assumed constant when designing the guidance curvature command. Furthermore, considering the time-varying missile speed, a novel robust geometric guidance law is designed in Ref. [14] and [15] by combining with the classical differential geometric curve theory and Lyapunov stability theory. In Ref. [16], the differential geometric guidance law is transformed from arc-length domain to time domain in 2D space, and the capture region for intercepting high-speed target is derived. In Ref. [17], the curvature and torsion of differential geometric guidance are further extended to 3D space. The PID control method is introduced in Ref. [18]. The high-frequency stability and robustness of the flight control system are verified through the classical frequency analysis method. In Ref. [19], the differential geometric guidance law under Frenet-Serret formula is re-derived in time domain, and its capture performance is verified. On this basis, the robust geometric guidance law is compared with traditional PN in Ref. [20] through simulation analysis, which shows that the robust geometric guidance law has stronger anti-disturb ability.

Inspired by the above references, the assumption of constant missile speed can be eliminated using the differential geometric theory to analyze the guidance law, which is a tremendous improvement comparing with traditional analysis methods in time domain. Besides, when intercepting a stationary target, the relative velocity vector between missile and target is the missile velocity vector. The command acceleration of PPN is also perpendicular to the missile target relative velocity, which indicates that PPN is equivalent to IPN. Then, the differential geometry theory is taken into consideration to establish the relative motion equation in arc-length domain. On this basis, the closed-form solutions of time-varying speed missile against stationary target with PPN are deduced by eliminating the influence of missile time-varying speed. Then, regarding the missile acceleration saturation, some important conclusions are obtained by analyzing the capture region and interception performance of time-varying speed missile against stationary target. Finally, the accuracy of these conclusions is verified by numerical simulation.

## 2   Problem Formulation

The 2D engagement geometry is shown in Fig. 1. $\boldsymbol{r}_\mathrm{m}$ and $\boldsymbol{r}_\mathrm{t}$ are the position vectors of missile and target in the 2D inertial coordinate frame $o\text{-}xy$, respectively. The LOS direction is the line from missile to target. The LOS angle $q$ is measured from refence line $ox$ to LOS, where the positive direction is counterclockwise. Similarly, $\varphi_\mathrm{m}$ and $\theta_\mathrm{m}$ are missile flight path angle and velocity leading angle respectively as shown in Fig. 1. The LOS coordinate frame consists of two unit-vectors $\boldsymbol{e}_r$ and $\boldsymbol{e}_\theta$, which are along and perpendicular to the LOS direction respectively. $\boldsymbol{t}_\mathrm{m}$ and $\boldsymbol{n}_\mathrm{m}$ are two unit-vectors along and perpendicular to missile velocity vector $v_m$.

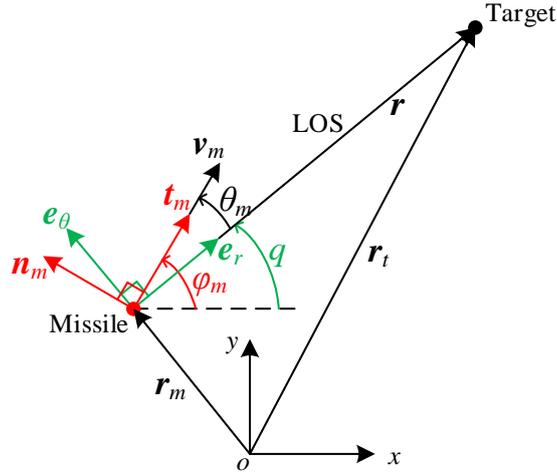

Fig. 1 2D engagement geometry

The 2D differential geometric curve theory is used to derive the relative motion equation of missile and target:

$$s_m = \int_0^t v_m(\sigma) d\sigma \quad \text{or} \quad \frac{ds_m}{dt} = v_m(t) \tag{1}$$

where $s_m$ is the arc-length of missile trajectory and $v_m(t)$ is the real-time speed of missile.

The relative distance vector $r$ between missile and target can be expressed as

$$r = r_t - r_m \tag{2}$$

Differentiating Eq. (2) with respect to the arc-length $s_m$, and expressing the relative velocity vector $r'$ with tangential component along $e_r$ and normal component along $e_\theta$, we have

$$r' = mt_t - t_m = r'e_r + rq'e_\theta \tag{3}$$

where $m = v_t/v_m$. $r'$ and $q'$ are the relative closing speed and LOS rate obtained by differentiating the relative distance vector $r$ with respect to the arc-length $s_m$. In addition, all the following variables with " $'$ " also donate the derivative with respect to $s_m$.

Expressing as components in $e_r$ and $e_\theta$, Eq. (3) is written as:

$$\begin{cases} r' = m\cos\theta_t - \cos\theta_m \\ rq' = m\sin\theta_t - \sin\theta_m \end{cases} \tag{4}$$

where the target leading angle $\theta_t$ is from $e_r$ to $t_t$, while the missile leading angle $\theta_m$ is from $e_r$ to $t_m$.

When the target is stationary, Eq. (4) can be simplified as:

$$\begin{cases} r' = -\cos\theta_m \\ rq' = -\sin\theta_m \end{cases} \tag{5}$$

Again, taking the derivative of Eq (3) with respect to $s_m$, and applying the Frenet-Serret formula, we have:

$$(r'' - rq'^2)\boldsymbol{e}_r + (rq'' + 2r'q')\boldsymbol{e}_\theta = m'\boldsymbol{t}_t + m^2 k_t \boldsymbol{n}_t - k_m \boldsymbol{n}_m \tag{6}$$

where $m' = (v_t/v_m)' = (v_t' v_m - v_t v_m')/v_m^2$.

Expressing as components in $\boldsymbol{e}_r$ and $\boldsymbol{e}_\theta$, Eq. (6) is rewritten as:

$$\begin{cases} r'' - rq'^2 = m'\cos\theta_t + m^2 k_t \sin\theta_t - k_m \sin\theta_m \\ rq'' + 2r'q' = m'\sin\theta_t - m^2 k_t \cos\theta_t + k_m \cos\theta_m \end{cases} \tag{7}$$

which is the relative kinematic equation between the missile and target in the arc-length domain.

Similarly, when the target is stationary, Eq. (7) can be simplified as:

$$\begin{cases} r'' - rq'^2 = -k_m(\boldsymbol{n}_m \cdot \boldsymbol{e}_r) = -k_m \sin\theta_m \\ rq'' + 2r'q' = -k_m(\boldsymbol{n}_m \cdot \boldsymbol{e}_\theta) = k_m \cos\theta_m \end{cases} \tag{8}$$

Furthermore, taking Eq. (5) into Eq. (8) yields:

$$\begin{cases} r'' - rq'^2 = -rq' k_m \\ rq'' + 2r'q' = r' k_m \end{cases} \tag{9}$$

Besides, 2D IPN in time domain proposed in Ref. [9] is given as follows:

$$\boldsymbol{a}_{IPN} = N\boldsymbol{v} \times \boldsymbol{\omega}_s = N\boldsymbol{v} \times \dot{q}\boldsymbol{e}_z \tag{10}$$

where $\boldsymbol{e}_z$ is a unit-vector perpendicular to the plane, satisfying right-hand rule with $\boldsymbol{e}_r$ and $\boldsymbol{e}_\theta$. $\boldsymbol{v}$ is relative velocity vector between missile and target. When the target to be intercepted is a stationary target, we have $v_t = 0$. That is to say, the relative velocity vector $\boldsymbol{v}$ coincides with the missile velocity vector $\boldsymbol{v}_m$. As a result, it can be considered that PPN and IPN are equivalent, namely:

$$\begin{aligned} \boldsymbol{a}_{PPN} &= \boldsymbol{a}_{IPN} = N\boldsymbol{v} \times \boldsymbol{\omega}_s = N(\dot{r}\boldsymbol{e}_r + r\dot{q}\boldsymbol{e}_\theta) \times (\dot{q}\boldsymbol{e}_z) \\ &= N\dot{q}(r\dot{q}\boldsymbol{e}_r - \dot{r}\boldsymbol{e}_\theta) \end{aligned} \tag{11}$$

The guidance command in arc-length domain is missile trajectory curvature $k_m$. According to the relevant theory of differential geometry, the relationship between the missile guidance curvature and guidance acceleration is as follows:

$$k_m = \frac{a_m}{v_m^2} = Nq' \tag{12}$$

Combining Eq. (11) and Eq. (12), the PPN in the arc-length domain is given as:

$$k_{PPN}\boldsymbol{n}_m = \frac{N\dot{q}(r\dot{q}\boldsymbol{e}_r - \dot{r}\boldsymbol{e}_\theta)}{v_m^2} = Nq'(rq'\boldsymbol{e}_r - r'\boldsymbol{e}_\theta) \tag{13}$$

## 3 Analysis of Time-varying Missile Intercepting Stationary Target with PPN

When the target is stationary, combining Eq. (9) and Eq. (12), the relative motion equation of missile and target guided by PPN in arc-length domain is given as:

$$\begin{cases} r'' - rq'^2 = -Nrq'^2 \\ rq'' + 2r'q' = Nr'q' \end{cases} \tag{14}$$

Firstly, taking the derivative of the second equation of Eq (5) with respect to $s_m$, and substituting the second equation of Eq (14), yields

$$\theta'_m \cos\theta_m = -r'q' - rq'' = (1-N)r'q' \tag{15}$$

Then, the first equation of Eq. (5) is substituted into Eq. (15) and simplified to obtain:

$$\theta'_m = (N-1)q' \tag{16}$$

As a result, Eq. (5) can be rewritten as:

$$\begin{cases} r' = -\cos\theta_m \\ r\theta'_m = (1-N)\sin\theta_m \end{cases} \tag{17}$$

### 3.1 Closed-Form Solution Analysis of PPN

From the second equation of Eq (14), we get

$$q' = q'_0 \left(\frac{r}{r_0}\right)^{N-2} \tag{18}$$

where $q'_0$ is the initial LOS rate in arc-length domain, and $r_0$ is the initial relative distance between missile and target.

Substituting Eq. (18) into the first equation of Eq. (14) yields:

$$r'' = \frac{dr'}{ds} = \frac{dr'}{dr}\frac{dr}{ds} = \frac{r'dr'}{dr} = \frac{(1-N)q'^2_0 r^{2N-3}}{r_0^{2(N-2)}} \Rightarrow r'dr' = \frac{(1-N)q'^2_0 r^{2N-3}}{r_0^{2(N-2)}} dr \tag{19}$$

By integrating both sides of Eq. (19), we have:

$$\int_{r'_0}^{r'} r'dr' = \int_{r_0}^{r} \frac{(1-N)q'^2_0 r^{2N-3}}{r_0^{2(N-2)}} dr \Rightarrow \frac{r'^2 - r'^2_0}{2} = \frac{(1-N)q'^2_0}{r_0^{2(N-2)}} \frac{r^{2(N-1)} - r_0^{2(N-1)}}{2(N-1)}$$
$$= \frac{r_0^2 q'^2_0 \left[r_0^{2(N-1)} - r^{2(N-1)}\right]}{2r_0^{2(N-1)}} = \frac{r_0^2 q'^2_0}{2}\left[1 - \left(\frac{r}{r_0}\right)^{2(N-1)}\right] \tag{20}$$

Accordingly, rearranging Eq. (20), the relation between $r'$ and $r$ can be expressed as:

$$r'^2 = r'^2_0 + r_0^2 q'^2_0 \left[1 - \left(\frac{r}{r_0}\right)^{2(N-1)}\right] = 1 - r_0^2 q'^2_0 \left(\frac{r}{r_0}\right)^{2(N-1)} \tag{21}$$

From Eq. (18) and Eq. (21), we obtain:

$$r'^2 + (rq')^2 = 1 - r_0^2 q_0'^2 \left(\frac{r}{r_0}\right)^{2(N-1)} + r^2 q_0'^2 \left(\frac{r}{r_0}\right)^{2(N-2)} = 1 \tag{22}$$

Eq. (22) shows that the guidance command generated by PPN does not affect the speed of the missile, but only changes the speed direction, which is consistent with the physical effect that the command acceleration of PPN is perpendicular to the speed. However, this does not mean that the speed of the missile remains unchanged, because there are other factors such as air drag and earth gravity that cause the change of missile speed.

By substituting Eq. (18) into Eq. (5), the relationship between $\theta_m$ and $r$ are as follows:

$$\sin\theta_m = -rq' = -rq_0'\left(\frac{r}{r_0}\right)^{(N-2)} = -r_0 q_0'\left(\frac{r}{r_0}\right)^{(N-1)} = \sin\theta_{m0}\left(\frac{r}{r_0}\right)^{(N-1)} \tag{23}$$

Eq. (23) describes the pattern of missile leading angle guided by PPN. It can be seen that when intercepting a stationary target, the missile leading angle $\theta_m$ will gradually converge to 0 with the decrease of the relative distance $r$ between missile and target, which means that the missile speed will gradually converge to the LOS direction.

Then, from Eq. (12) and Eq. (18), the pattern of the guidance curvature command are as follows:

$$k_m = Nq' = Nq_0'\left(\frac{r}{r_0}\right)^{N-2} \tag{24}$$

which indicates that with the decrease of the relative distance $r$ between missile and target, the guidance curvature command $k_m$ gradually decreases to 0. Due to the actual missile maneuvering acceleration limit, $N > 2$ is usually selected to avoid the increase of guidance curvature, which will be illustrated in the following simulation.

Fig. 2 and Fig. 3 provide the relationship between LOS rate, guidance curvature and relative distance, respectively. As shown in Fig. 2, it can be seen that when $1 < N < 2$, the LOS rate $q'$ tends to be positive infinity with the decrease of relative distance $r$. When $N = 2$, we have constant LOS rate with $q' = q_0'$. When $2 < N < 3$, $q'$ gradually converge to 0 with the decrease of $r$, but it changes slowly at the beginning of guidance and changes rapidly when $r$ is close to 0, which leads to the rapid change of guidance command. When $N = 3$, $q'$ converges linearly to 0 with the decrease of $r$. When $N > 3$, $q'$ also gradually converges to 0 with the decrease of $r$, but when $N$ is too large, it changes rapidly at the beginning of guidance, which also shows that the value of guidance gain $N$ is usually 3~5 in engineering. In Fig. 3, the tendency of guidance curvature is the same as that of LOS rate, because in the parameters of guidance curvature command in Eq. (12), the guidance gain $N$ is constant and the effect of missile speed is eliminated.

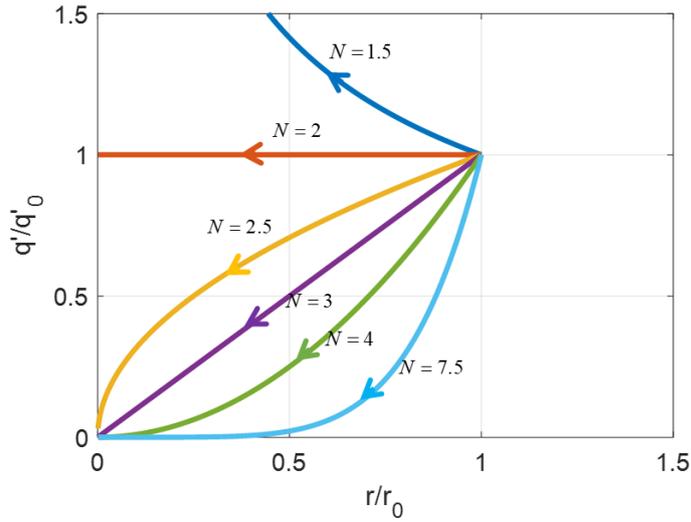

Fig. 2 The relationship between LOS rate and relative distance with different guidance gains

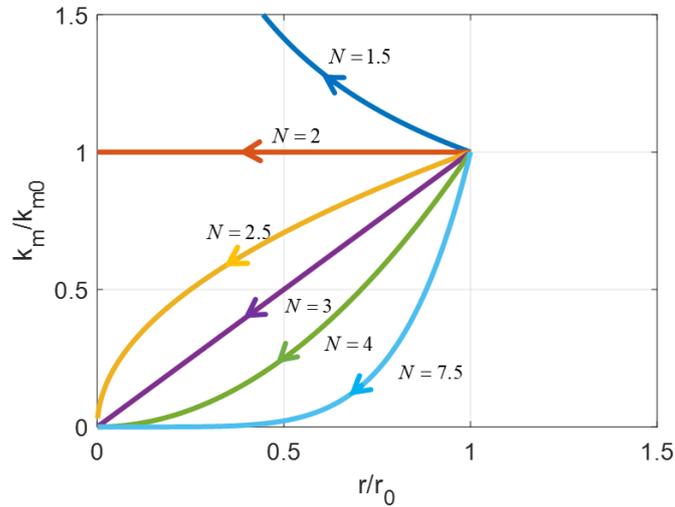

Fig. 3 The relationship between guidance curvature command and relative distance with different guidance gains

Fig. 4 presents the relationship between missile leading angle $\theta_m$ and relative distance $r$. It can be seen from Fig. 4 that with the increase of guidance gain $N$, the missile leading angle $\theta_m$ converges faster. When $1 < N < 2$, $\theta_m$ decreases slowly with the decrease of relative distance at first, and then decreases sharply as $r$ approaches 0. When $N > 2$, as $r$ approaches 0, the decreasing trend of $\theta_m$ gradually slows down.

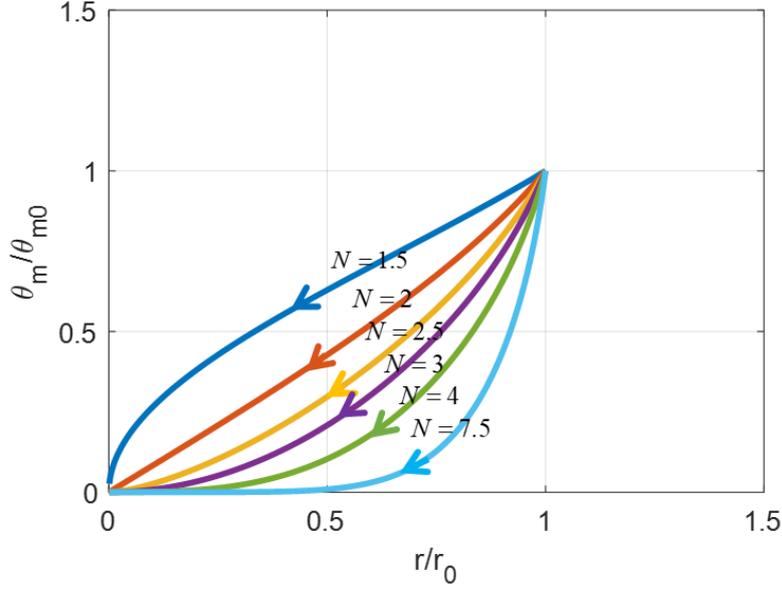

Fig. 4 The relationship between missile leading angle and relative distance with different guidance gains

### 3.2 Performance Analysis of PPN

During the process of missile attacking a stationary target, the main performance parameters concerned include: maximum maneuvering acceleration limit, energy consumption, flight time and capture region, et al. In this paper, the maximum maneuvering acceleration limit is equivalent to the maximum curvature, the energy consumption is described by curvature increment, and the flight time is equivalent to the flight path length.

When the leading angle is limited with $|\theta_m| \geq 90°$, we can easily conclude that the relative distance increases first and then decreases under the guidance of designed command from actual experiment. Now the theoretical analysis is presented. Considering Eq. (21), let $r' = 0$, the maximum relative distance between missile and target can be obtained:

$$r_{\max} = r_0 \left( \frac{1}{r_0 |q_0'|} \right)^{\frac{1}{N-1}} = r_0 \left( \frac{v_{m0}}{r_0 |\dot{q}_0|} \right)^{\frac{1}{N-1}} = r_0 \left( \sin |\theta_{m0}| \right)^{-\frac{1}{N-1}} \tag{25}$$

From Eq. (25), it can be seen that when the initial relative closing speed is greater than 0, the maximum relative distance between missile and target is only related to the initial parameters, and has nothing to do with the change of the missile speed during the whole interception process.

When $|\theta_m| < 90°$, guided by the designed command, the leading angle gradually converge to 0, while the relative distance also decreases. That is, the relative distance is the largest at the initial time. In summary, we get:

$$r_{max} = \begin{cases} r_0 (\sin|\theta_{m0}|)^{-\frac{1}{N-1}} &, 90° \leq |\theta_{m0}| < 180° \\ r_0 &, 0° \leq |\theta_{m0}| < 90° \end{cases} \quad (26)$$

It can be seen from Eq. (24) that when the relative distance is maximum, the missile guidance curvature $k_m$ takes the maximum value, while $N > 2$. By substituting the second equation of Eq. (5) and Eq. (26) into Eq. (24), the maximum guidance curvature can be obtained:

$$k_{mmax} = N|q_0'|\left(\frac{r_{max}}{r_0}\right)^{N-2} = \begin{cases} \frac{N}{r_0}(\sin|\theta_{m0}|)^{\frac{1}{N-1}} &, 90° \leq |\theta_{m0}| < 180 \\ \frac{N}{r_0}\sin|\theta_{m0}| &, 0° \leq |\theta_{m0}| < 90° \end{cases} \quad (27)$$

From Eq. (27), it is observed that the maximum guidance curvature is only related to the initial relative distance and the initial missile leading angle. The larger the initial relative distance, the smaller the maximum guidance curvature.

Also, the curvature increment required in the whole interception process, described by flight path angle increment $|\Delta\varphi|$, can be further calculated by Eq. (24):

$$|\Delta\varphi| = \int_{r_0}^0 |k_m(s)|ds = \int_{r_0}^0 N|q_0'|\left(\frac{r}{r_0}\right)^{N-2}ds = \frac{N|\theta_{m0}|}{(N-1)} \quad (28)$$

It should be noted that the curvature increment in arc-length domain is equivalent to the speed increment in time domain, which is used to describe energy consumption.

Substituting Eq. (23) into Eq. (17) in order to eliminate $r$, sorting and simplifying, we can get:

$$ds_m = -\frac{r_0}{(N-1)(\sin\theta_{m0})^{\frac{1}{N-1}}}(\sin\theta_m)^{-\frac{N-2}{N-1}}d\theta_m \quad (29)$$

Since Eq. (23) has presented that the missile leading angle gradually converges to 0 with the decrease of the relative distance, the flight path $S_m$ of the whole interception process can be calculated by integrating both sides of Eq. (29):

$$S_m = -\frac{r_0}{(N-1)(\sin\theta_{m0})^{\frac{1}{N-1}}}\int_{\theta_{m0}}^0 (\sin\theta_m)^{-\frac{N-2}{N-1}}d\theta_m \quad (30)$$

When the initial conditions of interception are known, the flight path of the whole interception process is determined. However, the Eq. (30) has an analytical solution only when $N = 2$, namely Eq. (34) below. When $N > 2$, the method of partial integration can be considered to calculate the series solution, or Taylor Expansion can be used to obtain the approximate solution. In addition, it also can be concluded from Eq. (30) that the flight trajectory is only related to the initial missile leading angle, initial relative distance and guidance gain, and has nothing to do with the initial missile speed and the speed change during the whole process of interception.

As presented in Fig. 5, when the initial missile leading angle is less than 90°, the maximum relative distance is the initial relative distance between missile and target. As a contrast, when the initial missile leading angle is greater than 90°, the maximum relative distance increases with the increase of the initial missile leading angle. Besides, the smaller $N$, the greater the maximum relative distance, which is consistent with the result of Eq. (26).

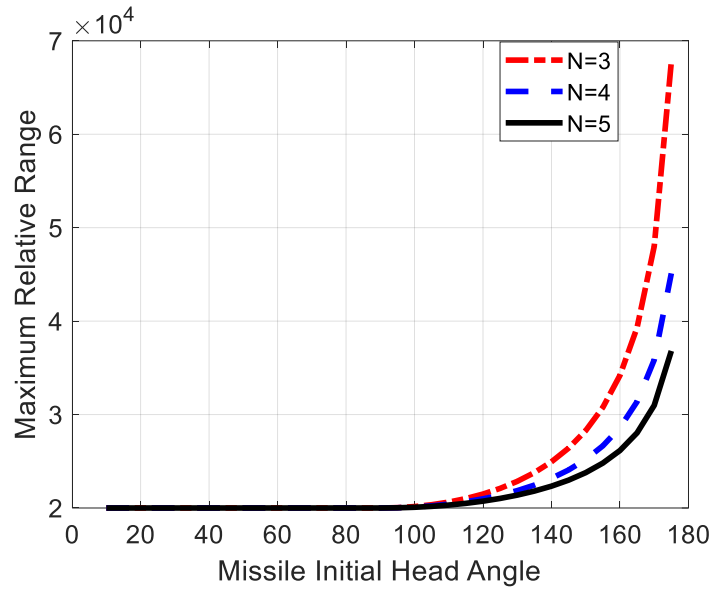

Fig. 5 The relationship between initial missile leading angle and maximum relative distance with different gains

Fig. 6 shows the relationship between initial missile leading angle and the maximum guidance curvature. With the increase of initial missile leading angle, the missile maximum guidance curvature first increases and then decreases. When the initial missile leading angle is constant, the greater the guidance gain coefficient $N$, the greater the maximum guidance curvature, which verifies the correctness of Eq. (27).

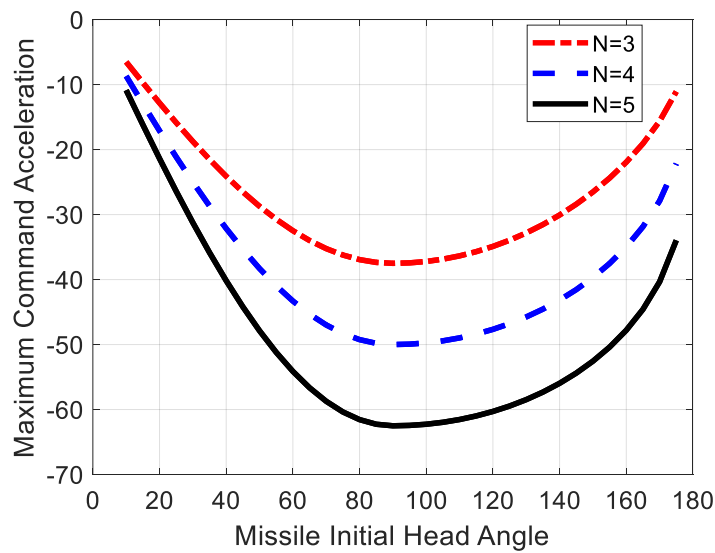

Fig. 6 The relationship between initial missile leading angle and maximum guidance curvature with different gains

Fig. 7 and Fig. 8 are the pattern of curvature increment and flight path with missile leading angle, respectively. From Fig. 7, it can be concluded that the curvature increment increases linearly with the leading angle of the missile's initial speed. The slope is $N/(N-1)$. The larger the guidance gain $N$, the smaller the slope, that is, the slower the curvature increment changes. As shown in Fig. 8, when the missile leading angle is small, the flight path increases slowly. As a contrast, when the missile leading angle is close to 180°, the flight path increases rapidly. Besides, the smaller the guidance gain $N$, the larger the flight path.

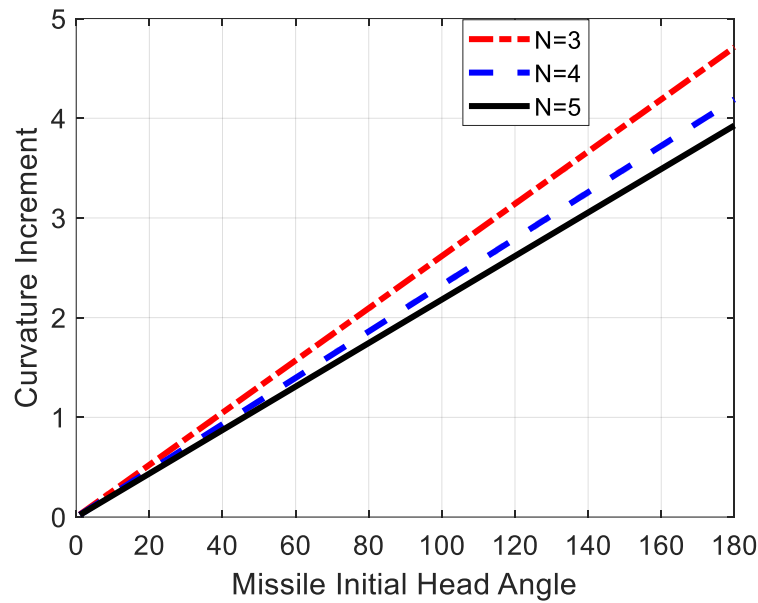

Fig. 7 The relationship between initial missile leading angle and curvature increment with different gains

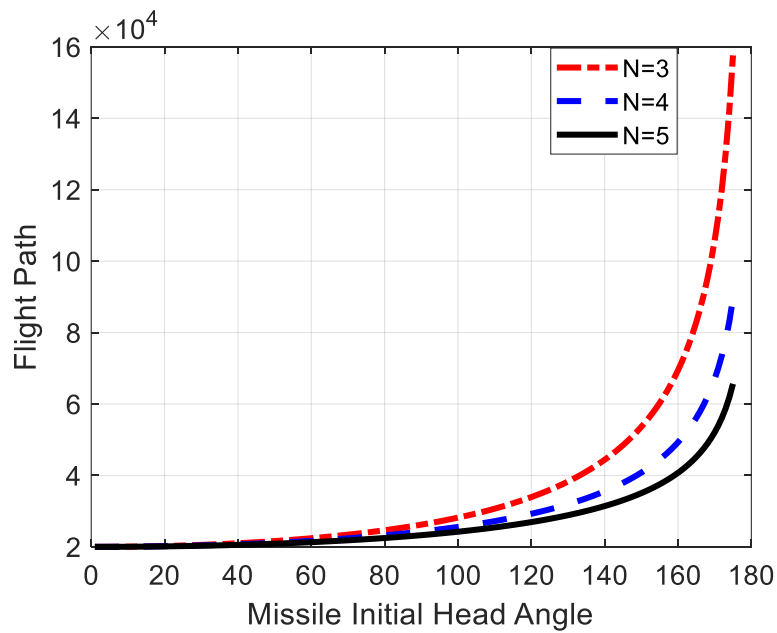

Fig. 8 The relationship between initial missile leading angle and flight path with different gains

### 3.3 Capture Region Analysis of PPN

From Eq. (23), we get:

$$r = r_0 \left( \frac{\sin \theta_m}{\sin \theta_{m0}} \right)^{\frac{1}{N-1}} \tag{31}$$

Since $r > 0$, $\sin \theta_m$ has the same sign as $\sin \theta_{m0}$. The following is divided into four cases for analysis.

Case 1: for the case where $\sin \theta_m > 0$ and $\sin \theta_{m0} > 0$, namely $0 < \theta_{m0} < \pi$, substituting Eq. (31) into the second equation of Eq. (17) yields:

$$\theta'_m = -\frac{(N-1)(\sin \theta_{m0})^{\frac{1}{N-1}}}{r_0} (\sin \theta_m)^{\frac{N-2}{N-1}} \tag{32}$$

From Eq. (32), for any $N > 1$, we have $\theta'_m < 0$, which indicates that $\theta_m$ will decrease monotonically, that is:

$$\theta_m \to 0 \quad ,\text{if} \quad 0 < \theta_{m0} < \pi \tag{33}$$

In particular, when $N = 2$, $\theta'_m$ is a constant. That is, $\theta_m$ will decrease to 0 linearly and flight path $S_m$ could be calculated as

$$S_m = \frac{r_0 \theta_{m0}}{(N-1)(\sin \theta_{m0})^{\frac{1}{N-1}}} \quad ,\text{if} \quad 0 < \theta_{m0} < \pi \tag{34}$$

When $N > 2$, as $\theta_m$ decreases to 0, $\theta'_m$ also gradually approaches 0, and as can be seen from Eq. (16), $q'$ also gradually approaches 0.

From Eq. (31), when $N > 1$, the relative distance $r$ will gradually decrease to 0 under the guided by the designed guidance law. Especially when $N = 2$, $r$ is linear with $\sin \theta_m$.

Case 2: for the case where $\sin \theta_m < 0$, namely $-\pi < \theta_m < 0$, the same analysis process is taken by replacing $\sin \theta_m$ with $-\sin \theta_m$ in Eq. (32).

Case 3: for the case where $\theta_m = \pi$, the missile is far away from the target along the LOS, and cannot intercept target successfully only by the guidance law adopted which is an unstable state in reality.

Case 4: for the case where $\theta_m = 0$, the missile approaches the target along the LOS. Theoretically, the missile can intercept target successfully without guidance command, that is, the guidance command is 0.

In summary, when the missile maneuvering acceleration limit is not considered, the capture region is $-180° < \theta_{m0} < 180°$. That is, as long as $N > 1$ and $|\theta_{m0}| \neq \pi$, the PPN can intercept stationary target theoretically.

When taking the missile's maneuvering acceleration limit into consideration, which is assumed as $\alpha_s$ in arc-length domain, the maximum guidance curvature should not be greater than the maximum maneuvering acceleration of the missile, namely:

$$k_{mmax} \leq \alpha_s \quad (35)$$

The maximum maneuvering acceleration of the missile described in time domain is converted into arc-length domain, which can be approximately expressed as:

$$\alpha_s \approx \frac{\alpha}{\max\{v_m^2(t)\}} \quad (36)$$

From Eq. (27), the maximum guidance curvature is only related to the initial relative distance and initial missile leading angle. And if $r_0$ remains constant, when $|\theta_{m0}| = \pi/2$, the maximum guidance curvature takes the maximum value. Moreover, if $\max\{k_{mmax}\} \leq \alpha_s$, the capture region is $-180° < \theta_{m0} < 180°$ which is same as the analysis from Case 1 to Case 4 mentioned. But if $\max\{k_{mmax}\} > \alpha_s$, then there are certain constraints on the value range of $\theta_{m0}$. At this time, substituting Eq. (27) into Eq. (35), and after calculation and simplification, the capture region of PPN against stationary target can be obtained:

$$\begin{cases} |\theta_{m0}| \geq \sin^{-1}\left(\left(\frac{r_0 \alpha_s}{N}\right)^{N-1}\right), & 90° \leq |\theta_{m0}| < 180 \\ |\theta_{m0}| \leq \sin^{-1}\left(\frac{r_0 \alpha_s}{N}\right), & 0° \leq |\theta_{m0}| < 90° \end{cases} \quad (37)$$

For the case where the missile's maneuvering acceleration is limited, in order to intercepting target successfully, when the absolute value of the initial missile leading angle is greater than or equal to 90°, the initial missile leading angle shall be greater than or equal to a certain value. As a contrast, when the absolute value of the initial missile leading angle is less than 90°, the initial missile leading angle shall be less than or equal to a certain value depicted in Eq. (37).

Fig. 9 provides the relationship between the capture region of PPN against stationary target and the initial relative distance when the initial missile speed and gain is known. In this case, the capture region of PPN increases with the increase of the initial relative distance. What's more, when the initial relative distance satisfies $r_0 \geq \sqrt{N/\alpha_s}$, the capture region is $-180° < \theta_{m0} < 180°$.

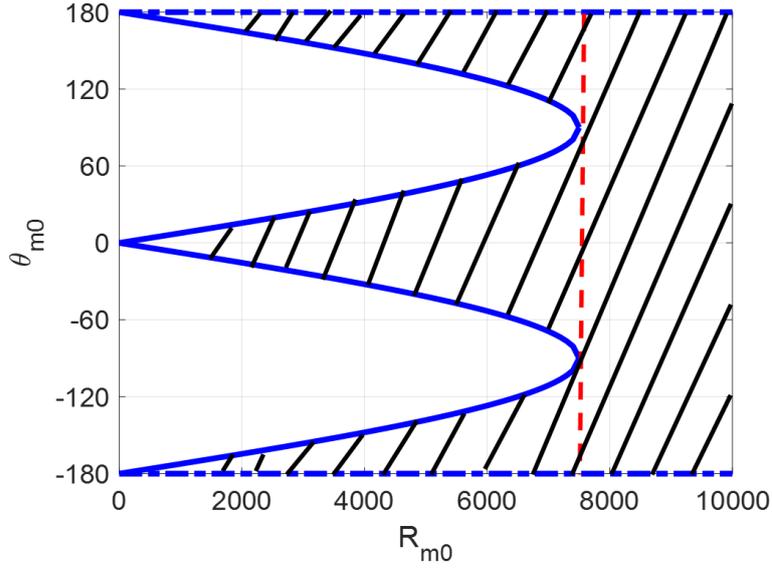

Fig. 9 Capture Region 1 ($v_{m0} = 500$, $N = 3$)

Similarly, the capture region shown in Fig. 10 is with known initial relative distance and guidance gain. In this case, when the initial relative distance and guidance gain is determined, if the initial speed is limited in $v_{m0} \leq \sqrt{r_0 \alpha / N}$, the capture region is $-180° < \theta_{m0} < 180°$. And with the increase of the initial speed, the range of the capture region decreases.

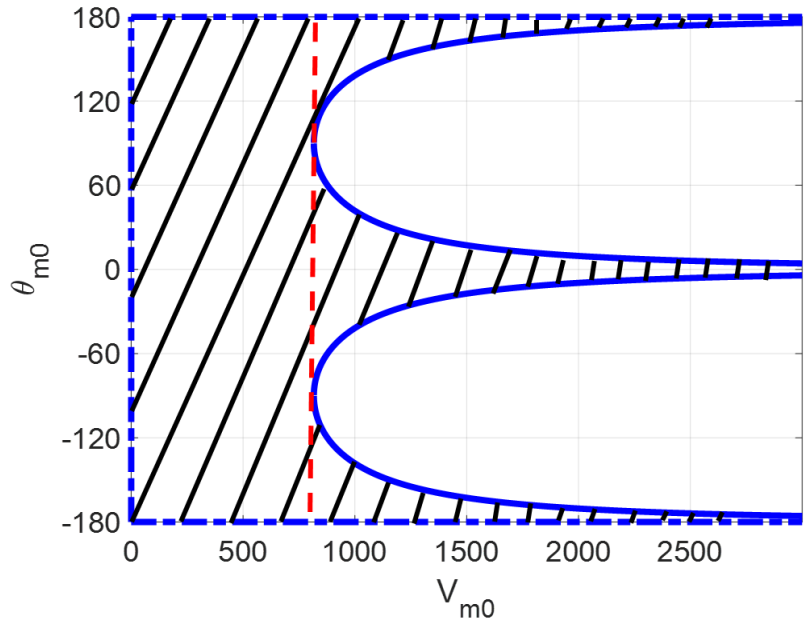

Fig. 10 Capture Region 2 ($r_0 = 20000$, $N = 3$)

## 4 Numerical Simulation

In this section, the conclusions derived in this paper are verified by numerical simulation. The initial states of missile and target in inertial coordinate frame is given in Table 1. Therefore, the initial relative distance is $r_0 = 20000\text{m}$, and the initial LOS angle is $q_0 = -60°$ from the table.

Table 1 Initial States of Missile and Target

| Quantity | Symbol | Value |
|---|---|---|
| Target initial position on x axis | $x_{t0}$ | 0m |
| Target initial position on y axis | $y_{t0}$ | 0m |
| Missile initial position on x axis | $x_{m0}$ | 10000m |
| Missile initial position on y axis | $y_{m0}$ | 17320m |
| Missile initial flight path angle | $\varphi_{m0}$ | 0° |
| Missile initial speed | $v_{m0}$ | 500m/s |

In addition, it is assumed that the acceleration caused by air drag on missile is $0.1\text{m/s}^2$ with the direction opposite to the missile velocity vector, that is, the air drag only changes the speed and does not change the direction of the speed. Besides, no measure error or dynamic lag is considered here, because the numerical simulation is mainly to verify the new closed-form solution in arc-length domain. In Scenario 1, the missile's initial leading angles are different with $N = 3$. In Scenario 2, the guidance gains are different with $\theta_{m0} = 120°$.

### 4.1 Scenario 1: Different Initial Leading Angle with $N = 3$

For Scenario 1 with $N = 3$, the initial leading angle of missile is chosen as $\theta_{m0} = \pm 30°, \pm 60°, 90°, 120°$ to demonstrate the new results in Section 3.

Fig. 11 shows the missile trajectories with different initial leading angle. As presented in Fig. 11, the larger the value of $|\theta_{m0}|$, the more curved the missile trajectory during the whole interception process, and the longer the flight path length. This is corresponding to Eq. (30). Besides, the terminal impact angles are also different corresponding to different initial leading angles, which is a potential application in the design of impact angle control guidance law.

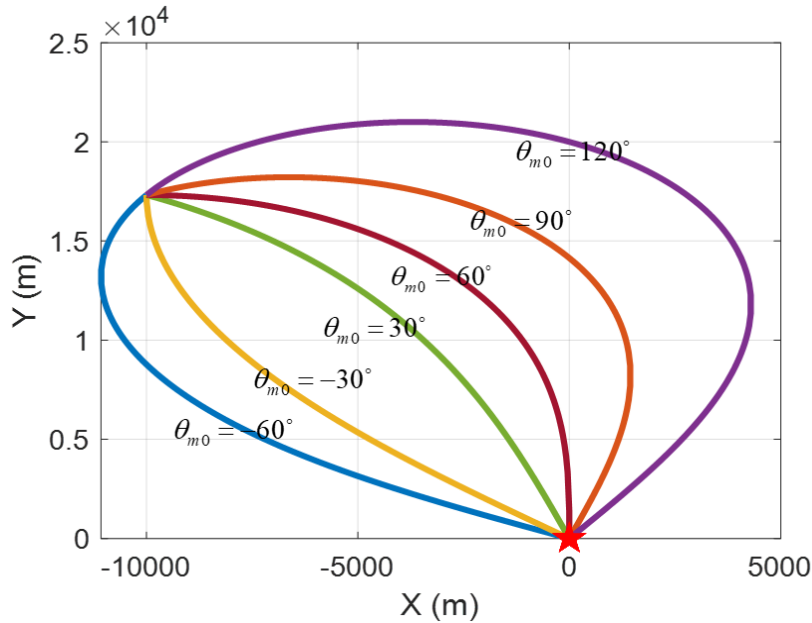

Fig. 11 Missile trajectories with different initial leading angles ($N = 3$)

The LOS rate and closing speed are presented in Fig. 12 and Fig. 13, from which it can be seen directly that the larger $|\theta_{m0}|$, the longer the interception time needed, which corresponds to the longer the missile path length during the interception process depicted in Fig. 11. As shown in Fig. 12, when $|\theta_{m0}| \leq 90°$, $|\dot{q}|$ converge to 0 gradually. And the larger $|\theta_{m0}|$, the slower the convergence rate of $|\dot{q}|$. As a contrast, when $|\theta_{m0}| = 120°$, $|\dot{q}|$ first increases and then decreases, and the initial value of $|\dot{q}|$ corresponding to $|\theta_{m0}|$ is the same as that of $|\dot{q}|$ corresponding to $180° - |\theta_{m0}|$. Besides, when $\theta_{m0}$ is a negative value, $\dot{q}$ has the opposite sign, equal size and same tendency comparing with that when the sign of $\theta_{m0}$ is positive. From Fig. 13, the value of closing speed $\dot{r}$ from missile to target converges to the value of the missile speed $v_m$ gradually if $|\theta_{m0}| \leq 90°$. As a contrast, the value of closing speed is positive, indicating that the missile is far away from the target if $|\theta_{m0}| = 120°$, which is consistent with reality. Then guided by PPN, the value of closing speed also decreases gradually, and finally converges to the value of the missile speed. Similarly, when $\theta_{m0}$ is a negative value, $\dot{r}$ has equal size and same tendency comparing with that when the sign of $\theta_{m0}$ is positive.

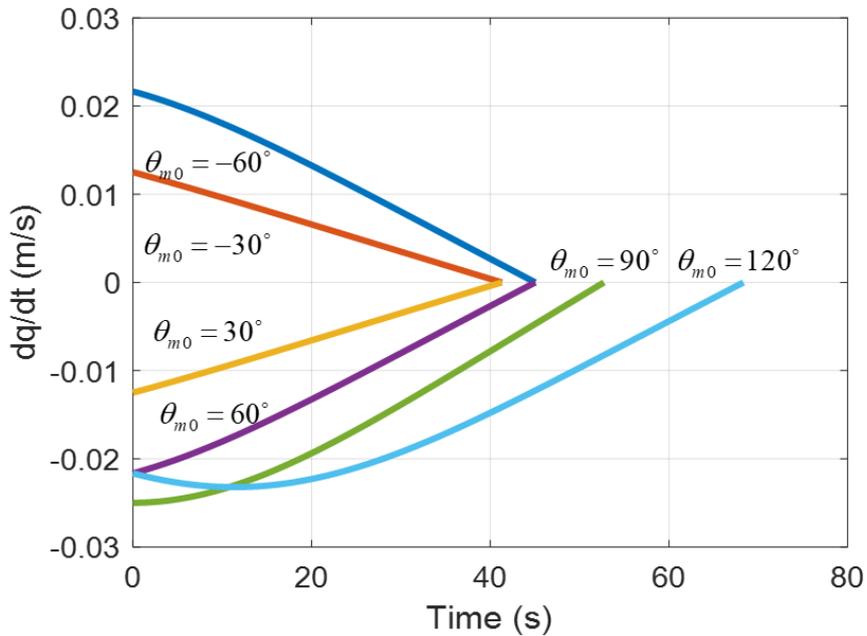

Fig. 12 LOS rate with different initial leading angles ($N = 3$)

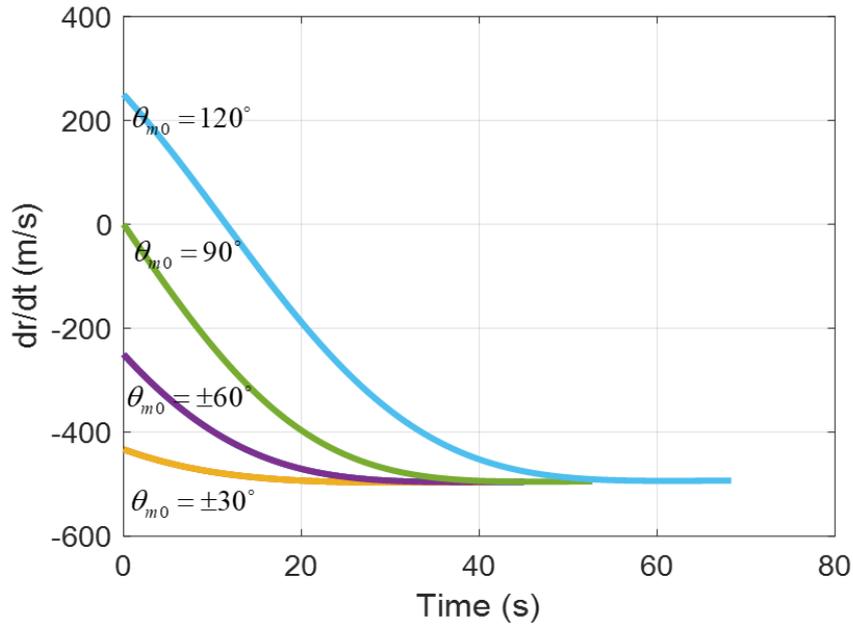

Fig. 13 Closing speed with different initial leading angles ($N = 3$)

Fig. 14 compares the relative distance with different initial leading angles under $N = 3$. When the guidance gain $N$ remains constant, the larger $|\theta_{m0}|$, the longer time is taken for the relative distance to converge to 0. And the relative distance first increases and then decreases if $|\theta_{m0}| = 120°$, which is consistent with Eq. (30). Fig. 15 shows the guidance curvature with time during interception, whose tendency is the same as that of LOS rate.

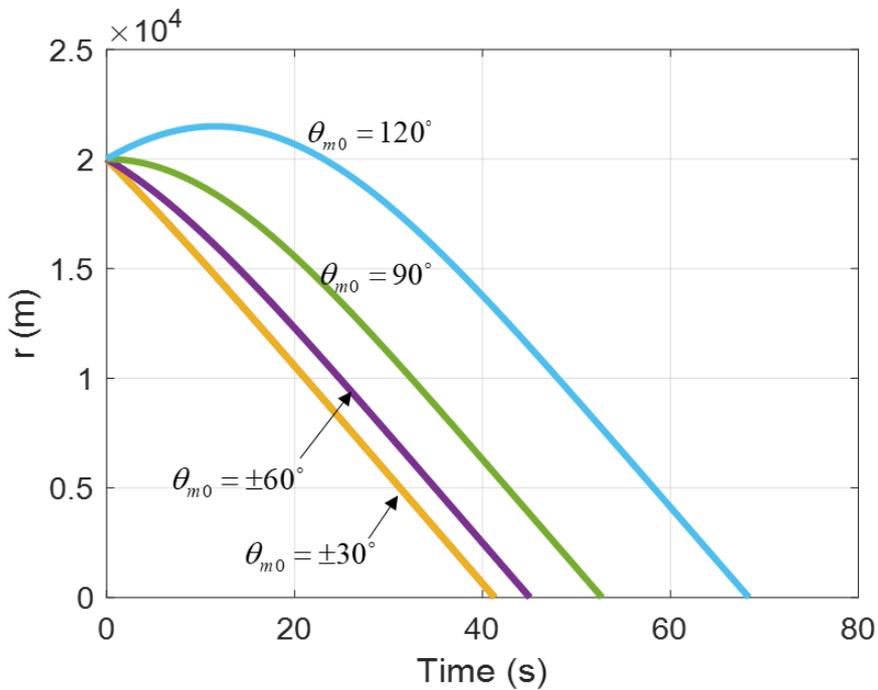

Fig. 14 Relative distance with different initial leading angles ($N = 3$)

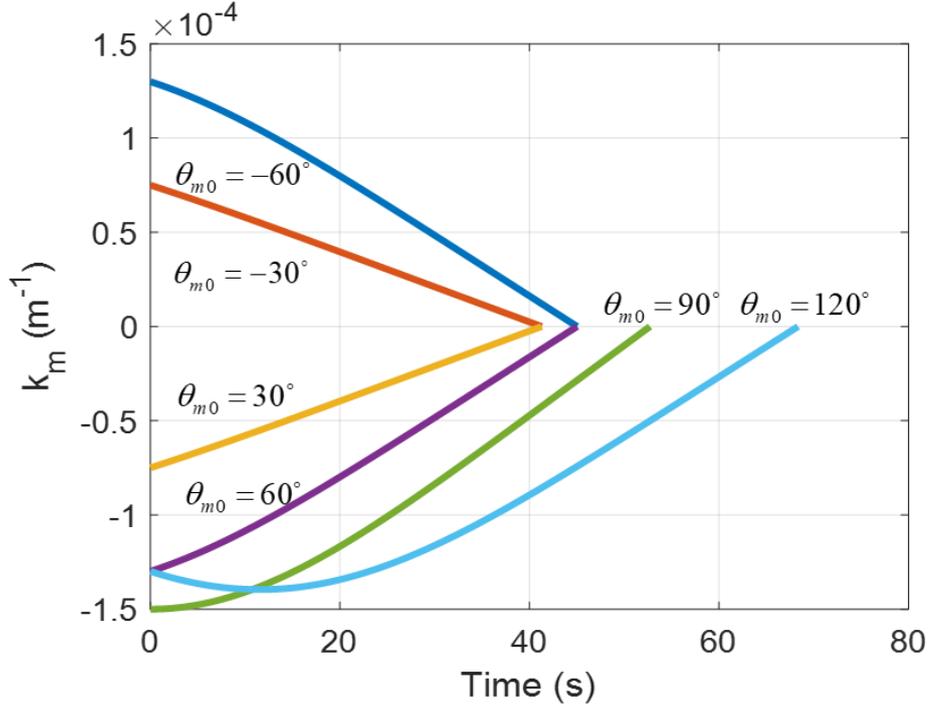

Fig. 15 Guidance curvature with different initial leading angles ($N = 3$)

Table 2 shows the theoretical and simulation results of flight path, curvature increment, and maximum relative distance with different initial leading angles. Although the missile speed is not constant, the error between the simulation results and theoretical results is very small and can be ignored, because the theoretical results of these variables are independent of the missile speed. Besides, it also can be concluded from Table 2 that the flight path increases gradually with the increase of $|\theta_{m0}|$, and do not be affected whether $\theta_{m0}$ takes a minus sign or not. Then, the curvature increment also increases with the increase of $|\theta_{m0}|$, and has a linear relationship as shown in Eq. (30). Last but not least, the maximum relative distance is the initial value $r_0$ if $|\theta_{m0}| < 90°$, while the maximum relative distance is larger than the initial value $r_0$ if $|\theta_{m0}| > 90°$, which is consistent with Eq. (26).

Table 2 Theoretical and simulation results with different initial leading angles

| Initial Leading Angle $\theta_0$ | | -60° | -30° | 30° | 60° | 90° | 120° |
|---|---|---|---|---|---|---|---|
| Fight Path (m) | Numerical | 22414.26 | 20561.14 | 20561.13 | 22414.26 | 26220.58 | 33937.42 |
| | Simulation | 22414.11 | 20560.25 | 20560.75 | 22414.11 | 26220.54 | 33936.96 |
| Curvature Increment | Analytical | 1.57080 | 0.785398 | 0.785398 | 1.57080 | 2.35619 | 3.14159 |
| | Simulation | 1.57083 | 0.785417 | 0.785417 | 1.57083 | 2.35623 | 3.14163 |
| Max Relative Distance (m) | Analytical | 20000 | 20000 | 20000 | 20000 | 20000 | 21491.40 |
| | Simulation | 20000 | 20000 | 20000 | 20000 | 20000 | 21491.40 |

### 4.2 Scenario 2: Different Guidance Gain with $\theta_0 = 120°$

For Scenario 2 with $\theta_{m0} = 120°$, according to the analysis in Section 3, the guidance gain is chosen as $N = 2, 3, 4, 5, 6$ to demonstrate the new results in Section 3.

Fig. 16 is the missile trajectories corresponding to different guidance gains when $\theta_{m0} = 120°$. From Fig. 16, the smaller $N$, the more curved the missile trajectory, which means the longer the flight path length is. As a result, the smaller the absolute value of the corresponding terminal impact angle is.

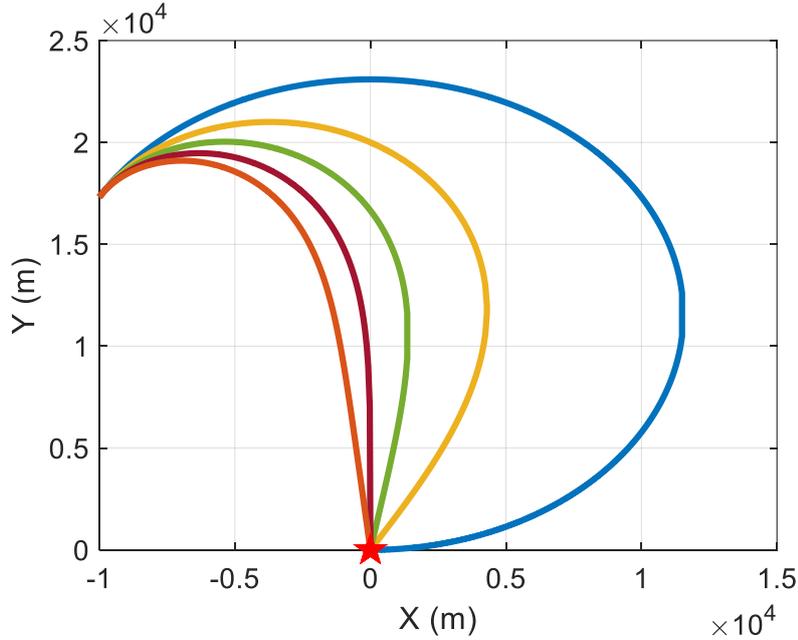

Fig. 16 Missile trajectories with different guidance gains ($\theta_{m0} = 120°$)

Fig. 17 and Fig. 18 present the curves of LOS rate and closing speed respectively. Similar to Scenario 1, the smaller $N$, the longer the time required for the interception, which corresponds to the longer flight path during the interception process shown in Fig. 16. As depicted in Fig. 17, $|\dot{q}|$ converge to 0 linearly if $N = 2$ while $|\dot{q}|$ first increases and then decreases if $N > 2$, which is consistent with Eq. (18). Besides, the larger $N$, the larger the maximum value of $|\dot{q}|$, the steeper the curve. It can be seen from Fig. 18 that the closing speed is positive at the initial time, indicating that the missile is far away from the target, then gradually decreases guided by PPN and converges to the value of missile speed, and the greater $N$, the steeper the curve.

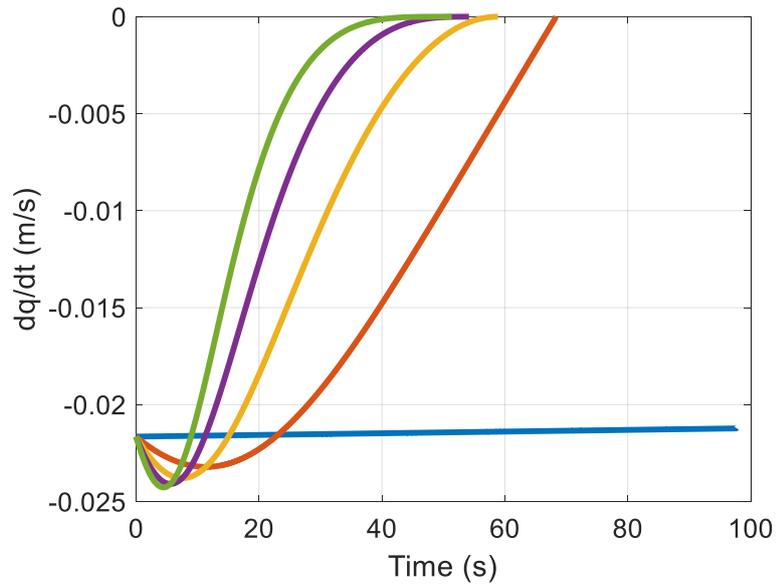

Fig. 17 LOS rate with different guidance gains ($\theta_{m0}$ =120°)

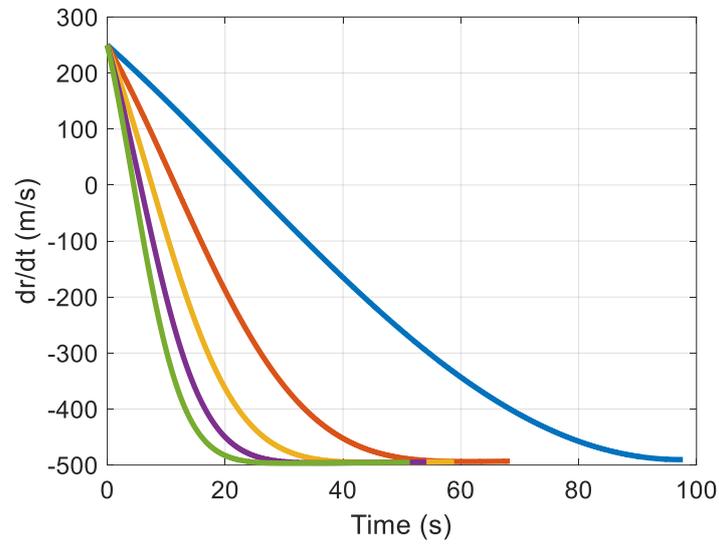

Fig. 18 Closing speed with different guidance gains ($\theta_{m0}$ =120°)

Fig. 19 shows the curve of guidance curvature during interception. The tendency of guidance curvature $k_m$ is the same as that of the LOS rate $\dot{q}$.

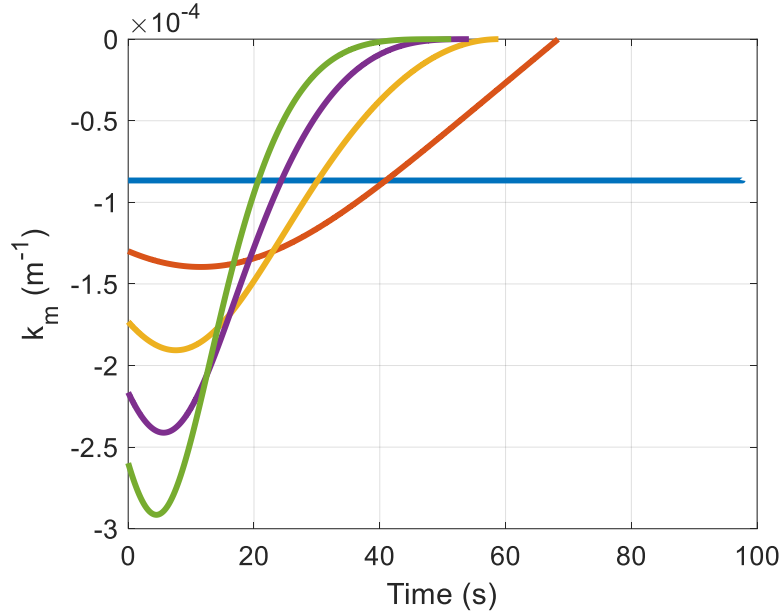

Fig. 19 Guidance curvature with different guidance gains ($\theta_{m0}$ =120°)

Table 3 shows the theoretical results and simulation results of flight path, curvature increment, and maximum relative distance corresponding to different guidance gains under $\theta_0 = 120°$. Although the missile speed changes, the error between the simulation results and theoretical results is very small and can be ignored. That is because the theoretical results of flight path, curvature increment, and maximum relative distance are independent of the missile speed. It can be seen from the table that with the increase of guidance gain $N$, the three variables mentioned gradually decrease, which is consistent with the theoretical analysis in Section 3.2.

Table 3 Theoretical and simulation results with different guidance gains

| Guidance Gain $N$ | | 2 | 3 | 4 | 5 | 6 |
|---|---|---|---|---|---|---|
| Fight Path (m) | Numerical | 48367.98 | 33937.42 | 29259.56 | 26936.62 | 25546.55 |
| | Simulation | 48367.83 | 33936.98 | 29259.25 | 26936.30 | 25546.13 |
| Curvature Increment | Analytical | 4.18879 | 3.14159 | 2.79253 | 2.61799 | 2.51327 |
| | Simulation | 4.18886 | 3.14163 | 2.79257 | 2.61805 | 2.51334 |
| Max Relative Distance (m) | Analytical | 23094.01 | 21491.40 | 20982.30 | 20732.29 | 20583.72 |
| | Simulation | 23094.01 | 21491.40 | 2.0982.30 | 20732.29 | 20583.72 |

## 5 Conclusion

In this paper, some new results on interception of stationary targets at arbitrary time-varying speed by PPN are obtained using differential geometric curve theory. In the design of guidance law, the hypothesis of constant missile speed in time domain is removed in arc-length domain. As a result, the closed-form solution of time-varying speed missile intercepting stationary target with PPN is obtained. And the guidance performance is analyzed. Then, considering the missile maneuvering acceleration limit, the capture region of time-varying speed missile intercepting stationary target is analyzed. Although it looks like the conclusion in time domain in form, the elimination of the constant speed hypothesis is significant progress. On the basis of these work, the novel guidance design and stability analysis of missile with time-varying speed in arc-length domain will be performed in future studies.